\documentstyle[amssymb,aps,epsfig]{revtex}
\begin{document}
\title{Getting More from Pushing Less: Negative Specific Heat and Conductivity in
Non-equilibrium Steady States}
\author{R. K. P. Zia$^{1,2}$, E. L. Praestgaard$^3$ and O. G. Mouritsen$^4$}
\address{$^1$Center for Stochastic Processes in Science and Engineering\\
Department of Physics \\
Virginia Polytechnic Institute and State University, Blacksburg, VA\\
24061-0435, USA \\
$^2$Fachbereich Physik, Universit\"{a}t - Gesamthochschule Essen,\\
D-45117 Essen, Federal Republic of Germany.\\
$^3$Department of Life Sciences and Chemistry, Roskilde University \\
4000 Roskilde, Denmark\\
$^4$Physics Department, University of Southern Denmark-Odense \\
Campusvej 55, DK-5230 Odense M, Denmark}
\date{8/8/01}
\maketitle

\begin{abstract}
\begin{center}
{\bf ABSTRACT}
\end{center}

For students familiar with equilibrium statistical mechanics, the notion of
a positive specific heat, being intimately related to the idea of stability,
is both intuitively reasonable and mathematically provable. However, for
system in {\em non-equilibrium} stationary states, coupled to more than one
energy reservoir (e.g., thermal bath), {\em negative} specific heat is
entirely possible. In this paper, we present a ``minimal'' system displaying
this phenomenon. Being in contact with two thermal baths at different
temperatures, the (internal) energy of this system may {\em increase} when a
thermostat is turned down. In another context, a similar phenomenon is {\em %
negative conductivity}, where a current may increase by decreasing the drive
(e.g., an external electric field). The counter-intuitive behavior in both
processes may be described as `` getting more from pushing less.'' The
crucial ingredients for this phenomenon and the elements needed for a
``minimal'' system are also presented.
\end{abstract}

\section{Introduction}

To most physicists, it is intuitively reasonable that the average internal
energy, $U$, of a system in contact with a thermal bath should increase with 
$T$, the bath temperature. For those well versed with equilibrium
statistical mechanics, it is trivial to prove this statement, usually in the
form of the specific heat, $\partial U/\partial T$, being positive. Indeed,
this notion is so deeply rooted in the idea of stability that its contrary
may be rejected immediately, even for {\em non-equilibrium} steady states.
In this paper, we show the presence of a {\em negative} specific heat in a
minimal system, coupled to two thermal baths at different temperatures. We
also point out a crucial ingredient which makes this phenomenon possible,
namely, the presence of ``barriers'' which trap the system in a higher
energy state. In a similar vein, we discuss {\em negative conductivity},
using again a simple model to demonstrate the phenomenon of increasing
current by decreasing the drive (e.g., an external electric field). Here
too, a crucial ingredient is the presence of ``obstructions'' which trap the
system in a low current state. Noting that both processes involve a similar
idea, we coin a ``folksy'' phrase -- getting more from pushing less -- to
describe this class of counter-intuitive behavior.

We begin with a brief summary of standard equilibrium statistical mechanics,
for which no dynamics is needed. In contrast, for {\em non-equilibrium}
steady states, the underlying dynamics (which leads a system into such
states) is crucial. The most convenient framework for describing both
scenarios is the master equation , which we review in the next section, for
the convenience of those readers unfamiliar with this approach. In this
context, a clear distinction between equilibrium and non-equilibrium steady
states will be presented. In section 3, we display the presence of negative
specific heat in some minimal systems and a more typical, many-body{\em \ }%
system, which are coupled to two thermal baths. Similarly, negative
conductivity is shown to arise in a simple hopping model in Section 4.
Concluding in the last section, we recapitulate the key ingredients needed
for such phenomena and draw an intuitive picture of the mechanism. For ease
of reading, the details of the algebra are taken out of the main text and
placed in an appendix.

\section{Contrasts between Equilibrium and Non-equilibrium Steady States}

To describe a physical system statistically, a complete and {\em relevant }%
set of microstates (or configurations: ${\cal C}$) for the system must first
be specified. For simplicity, let us consider finite systems with finite
degrees of freedom (e.g., finite Ising models), so that the ${\cal C}$'s
form a discrete, countable set. With appropriate mathematical tools, our
considerations can be extended to the thermodynamic limit of systems with
continuous degrees of freedom. Next, we need $P[{\cal C]}$, the probability
for finding the system in each ${\cal C}$. From here, the average of an
observable quantity (e.g., internal energy), which takes the value ${\cal O}[%
{\cal C}]$ in configuration ${\cal C}$, is then given by 
\begin{equation}
\left\langle {\cal O}\right\rangle =\sum_{{\cal C}}{\cal O}[{\cal C}]P[{\cal %
C]}\,.
\end{equation}
Beyond averages, fluctuations and correlations are similarly computed.

When a system is evolving, then the probability distribution is a function
of time: $P[{\cal C};t],$ leading to time dependent averages, fluctuations,
etc. Clearly, an ``equation of motion'' is needed to describe the evolution
of $P[{\cal C};t]$. In principle, we should derive such an equation from,
e.g., Newton's equations. In practice, however, this is typically not
feasible. Instead, let us follow the standard framework of a master equation 
\cite{Reichl}: 
\begin{equation}
\frac \partial {\partial t}P[{\cal C};t]=\sum_{{\cal C}^{\prime }}\left\{
R\left( {\cal C}^{\prime }\rightarrow {\cal C}\right) P[{\cal C}^{\prime
};t]-R\left( {\cal C}\rightarrow {\cal C}^{\prime }\right) P[{\cal C}%
;t]\right\}  \label{MEq}
\end{equation}
which is basically a continuity equation for the probability density. Here, $%
R\left( {\cal C}\rightarrow {\cal C}^{\prime }\right) $ stands for the rate
a configuration ${\cal C}$ changes to ${\cal C}^{\prime }$. Most frequently,
these rates are {\em not} time dependent, i.e., the dynamics is invariant
under time-translation. As in the case of Eq. (\ref{MEq}) itself, they can
be found in principle, once we specify how our system is coupled to its
environment (e.g., a thermal reservoir). Again, in practice, the task of
finding these rates is prohibitively complex, so that progress in this
approach relies, typically, on postulating reasonable forms, based on sound
physical principles. Returning to Eq. (\ref{MEq}), we see that, being
linear, the right hand side may be written as a matrix, ${\Bbb L}$,
operating on a ``vector.'' In short hand (with $\left| P\right\rangle _t=P[%
{\cal C};t]$), this equation takes the form 
\begin{equation}
\partial _t\left| P\right\rangle _t={\Bbb L}\left| P\right\rangle _t\,.
\label{MEq-sh}
\end{equation}
Explicitly, the matrix elements are 
\begin{equation}
{\Bbb L}_{{\cal C},{\cal C}^{\prime }}=R\left( {\cal C}^{\prime }\rightarrow 
{\cal C}\right) -\delta \left( {\cal C},{\cal C}^{\prime }\right) \sum_{%
{\cal C}^{\prime \prime }}R\left( {\cal C}\rightarrow {\cal C}^{\prime
\prime }\right)  \label{L}
\end{equation}
where $\delta \left( {\cal C},{\cal C}^{\prime }\right) $ is the Kronecker
delta.

For simplicity, let us focus on systems which eventually settle down into a
unique time-independent state, i.e., 
\begin{equation}
\lim_{t\rightarrow \infty }P({\cal C},t)=P^{*}({\cal C})\,.
\end{equation}
In other words, $P^{*}$ satisfies $\partial P^{*}/\partial t=0$ and is the
eigenvector of ${\Bbb L}$ with zero eigenvalue. Further, the real parts of
all other eigenvalues are strictly negative, so that $P^{*}$ is a {\em stable%
} state.

Clearly, this approach should embrace systems which evolve towards thermal
equilibrium. If we wish to reproduce the results from the theory of {\em %
equilibrium} statistical mechanics (e.g., the Boltzmann factor in the
canonical ensemble), there must be some constraints on these rates. These
constraints are known as {\em detailed balance}. Specifically, denoting the
energy of a configuration of our system be given by the Hamiltonian ${\cal %
H[C]}$, we must impose, for an isolated system with total energy $E$, 
\begin{equation}
R\left( {\cal C}^{\prime }\rightarrow {\cal C}\right) =R\left( {\cal C}%
\rightarrow {\cal C}^{\prime }\right)   \label{DB0}
\end{equation}
if ${\cal H[C}^{\prime }{\cal ]}={\cal H[C]}=E$ and $R\left( {\cal C}%
^{\prime }\rightarrow {\cal C}\right) =0$ otherwise. Then $P^{*}({\cal C}%
)\propto 1$, which is the fundamental hypothesis (microcanonical ensemble),
clearly satisfies ${\Bbb L}\left| P^{*}\right\rangle =0$. Generalizing to
the canonical case (with $\beta \equiv 1/k_BT$), we demand 
\begin{equation}
\frac{R\left( {\cal C}^{\prime }\rightarrow {\cal C}\right) }{R\left( {\cal C%
}\rightarrow {\cal C}^{\prime }\right) }=\exp \left\{ \beta \left( {\cal H[C}%
^{\prime }{\cal ]}-{\cal H[C]}\right) \right\}   \label{DB1}
\end{equation}
and verify that a distribution given by the Boltzmann factor 
\begin{equation}
P_{eq}^{*}({\cal C})\propto e^{-\beta {\cal H[C]}}\,
\end{equation}
indeed satisfies ${\Bbb L}\left| P_{eq}^{*}\right\rangle =0$. Apart from
these constraints, there is much leeway in postulating the $R$'s,
corresponding to the fact that equilibrium states are independent of the
details of the dynamics. In particular, it is not crucial that $R\left( 
{\cal C}\rightarrow {\cal C}^{\prime }\right) \neq 0$ for {\em all} pairs of 
$\left( {\cal C},{\cal C}^{\prime }\right) $, though there must be enough
non-zero rates so that every configuration (within the desired ensemble) may
be reached from any other one. As an explicit example, we give the
Metropolis rate\cite{metro}: 
\begin{equation}
R\left( {\cal C}^{\prime }\rightarrow {\cal C}\right) =\min [1,e^{\beta (%
{\cal H[C}^{\prime }{\cal ]}-{\cal H[C]})}]\,,  \label{metro}
\end{equation}
which is the basis of numerous successful Monte Carlo simulations of systems
in equilibrium.

From the condition of detailed balance, a stronger statement about an
equilibrium state emerges. Not only is the distribution time-independent,
the {\em net} (probability) current between{\em \ any} pair $\left( {\cal C},%
{\cal C}^{\prime }\right) $ vanishes! To be more explicit, we regard the
right hand side of the master equation (\ref{MEq}) as a sum of the net
currents (from ${\cal C}^{\prime }$ into ${\cal C}$ ): 
\begin{equation}
R\left( {\cal C}^{\prime }\rightarrow {\cal C}\right) P[{\cal C}^{\prime
};t]-R\left( {\cal C}\rightarrow {\cal C}^{\prime }\right) P[{\cal C};t].
\end{equation}
Then detailed balance, embodied in Eqs. (\ref{DB0}) or (\ref{DB1}), implies
the vanishing of {\em all }net currents. An analog of this situation in
electrodynamics is electro{\em statics}, where the charge distribution is
stationary and no currents exist.

Next, let us turn to {\em non-equilibrium }steady states. The simplest
example is a system in contact with {\em two} energy reservoirs, e.g., two
thermal baths at different temperatures: $T_1$ and $T_2$. For physically
realizable cases, we can think of a time frame during which energy flows 
{\em through} our system (from the hotter to the cooler bath) steadily, so
that the average energy {\em within} the system is time-independent.
Typically, to model the coupling of our system to such reservoirs, there is
no need to respect detailed balance when specifying a set of rates. Of
course, the rates are not completely free of contraints. Without going into
details, let us simply restrict our considerations to those rates which
eventually take the system to a unique time-independent distribution: $P^{*}(%
{\cal C})$. However, without detailed balance, we should expect,
generically, $P^{*}({\cal C})\neq P_{eq}^{*}({\cal C})$. Now, a
time-independent distribution does not imply that {\em all }currents are
zero. Instead, we can expect (time independent) current {\em loops} to be
present. Using the analog above, this situation corresponds to {\em %
magnetostatics}, in which the charge density is stationary but steady
current loops prevail. In this sense, we prefer to use the term ``steady
states'' for describing such (magnetostatic-like) systems, while reserving
``stationary states'' for those in equilibrium (or electrostatics).
The presence of current loops also highlights another crucial
difference between equilibrium and non-equilibrium steady states, namely,
time reversal invariance. Since currents change sign under time reversal,
only states with {\em zero} currents are invariant. In this light, the
concepts of equilibrium, detailed balance, and time reversal are intimately connected.

Once $P^{*}({\cal C})$ is known, we can compute the ``internal energy'' of
the system 
\begin{equation}
U\equiv \left\langle {\cal H[C]}\right\rangle \equiv \sum_{{\cal C}}{\cal H}[%
{\cal C}]P^{*}[{\cal C]}\,,
\end{equation}
since we have a well defined microscopic Hamiltonian. For equilibrium states
with $P_{eq}^{*}({\cal C})=e^{-\beta {\cal H[C]}}/\sum_{{\cal C}^{\prime
}}e^{-\beta {\cal H[C}^{\prime }{\cal ]}}$, it is a simple step to check
that, without knowing the details of ${\cal H[C]}$, the specific heat: $\partial U/\partial T\propto -\partial U/\partial \beta \propto \left\langle
\left( {\cal H[C]}-U\right) ^2\right\rangle $ is never negative. Notice that 
$\partial U/\partial T\geq 0$ is true for systems of {\em any} size,
regardless of whether a proper thermodynamic limit exists or not. For
non-equilibrium steady states in, e.g., systems coupled to two baths, we
naturally have $U(T_1,T_2)$ and may extend the usual definition of specific
heat. Of course, in this case, there are two such response functions: 
\begin{equation}
C_1\equiv \frac{\partial U}{\partial T_1}\quad \text{and}\quad C_2\equiv 
\frac{\partial U}{\partial T_2}
\end{equation}
as we can separately ``dial up'' the temperature of either bath while
keeping the other fixed. While we cannot expect {\em both} to be negative,
we will show that one of them can be {\em negative} in very simple systems.

\section{Simple Systems with Negative Specific Heat}

\subsection{A minimal system with three states}

We begin by considering an abstract system with only three, non-degenerate,
energy levels. For reasons shown at the end of the next subsection, it is
impossible to construct a system with only two microstates which displays
negative specific heat. In this sense, we believe that the three state
system is ``minimal.''

Each configuration (or microstate) is associated with a unique energy. Using
the subscript $\alpha =0,1,2$, let us denote the energies by 
\begin{equation}
E_\alpha =0,\varepsilon _1,\varepsilon _2
\end{equation}
with $\varepsilon _2>\varepsilon _1>0$. Our goal is, given a set of rates $%
R(\alpha \rightarrow \alpha ^{\prime })$, to find $U$, the average energy in
the steady state. First, we must compute $P_\alpha ^{*}$, the probability
for finding the system in level $\alpha $ when steady state has been
reached. Then 
\begin{equation}
U\equiv \sum_\alpha E_\alpha P_\alpha ^{*}=\varepsilon _1P_1^{*}+\varepsilon
_2P_2^{*}\,.
\end{equation}
Had we been interested in the equilibrium distribution, corresponding to our
system being in contact with a single thermal bath, then we could use, e.g.,
the usual Metropolis rates (Eq. \ref{metro}). Here, they are explicitly 
\begin{eqnarray}
R(0 &\rightarrow &1)=e^{-\beta \varepsilon _1}\,;\quad R(1\rightarrow 0)=1 
\nonumber \\
R(0 &\rightarrow &2)=e^{-\beta \varepsilon _2}\,;\quad R(2\rightarrow 0)=1 \\
R(1 &\rightarrow &2)=e^{-\beta \delta }\,\,\,;\quad R(2\rightarrow 1)=1 
\nonumber
\end{eqnarray}
where $\delta \equiv \varepsilon _2-\varepsilon _1$. The reader may verify
that, with these rates in Eq. (\ref{L}), the Boltzmann form, $%
P_{1,2}^{*}=P_0^{*}\exp \left( -\beta \varepsilon _\alpha \right) $, is
indeed a solution to ${\Bbb L}\left| P^{*}\right\rangle =0$. Note further
that, even if we forbid the $0\leftrightarrow 1$ transition (by setting $%
R(0\rightarrow 1)=R(1\rightarrow 0)=0$), the system will still equilibrate
to the same set of $P^{*}$'s.

Next, let us modify the dynamics by coupling the $0\leftrightarrow 2$ and $%
1\leftrightarrow 2$ transitions to {\em different }baths, while forbidding
the $0\leftrightarrow 1$ transition entirely. To avoid confusion with
subscripts, we will label the two bath temperatures by $T_x$ and $T_y.$
Defining 
\begin{equation}
x\equiv \exp \left[ -\varepsilon _2//k_BT_x\right] \,\quad \text{and}\quad
y\equiv \exp \left[ -\delta /k_BT_y\right] \,
\end{equation}
the master equation takes the simple form 
\begin{eqnarray}
\partial _tP_0 &=&-xP_0+P_2  \nonumber \\
\partial _tP_1 &=&-yP_1+P_2 \\
\partial _tP_2 &=&xP_0+yP_1-2P_2  \nonumber
\end{eqnarray}
The steady state distribution is trivial to find: 
\begin{equation}
P_0^{*}=y/Z\,;\,P_1^{*}=x/Z\,;\,\,P_2^{*}=xy/Z\,\,,
\end{equation}
where $Z\equiv $ $x+y+xy$. Thus, the average energy is 
\begin{equation}
U=\frac{x\varepsilon _1+xy\varepsilon _2}{x+y+xy}
\end{equation}
so that the specific heats (associated with the $x$- and $y$-baths) are: 
\begin{equation}
C_x\equiv \frac{\partial U}{\partial T_x}=\left\{ \varepsilon
_1+y\varepsilon _2\right\} \frac{xy\varepsilon _2}{k_BT_x^2\left[
x+y+xy\right] ^2}\,
\end{equation}
and 
\begin{equation}
C_y\equiv \frac{\partial U}{\partial T_y}=\left\{ x\varepsilon
_2-\varepsilon _1\left( 1+x\right) \right\} \frac{xy\delta }{k_BT_y^2\left[
x+y+xy\right] ^2}\,\,.  \label{Cy}
\end{equation}
While the first of these never goes negative, the second clearly becomes 
{\em negative} for a range of $x$. Working out the details, we find $C_y<0$ (%
{\em for all} $T_y$!), provided $T_x$ drops below the critical value 
\begin{equation}
T_{xc}\equiv \frac{\varepsilon _2}{k_B}\left[ \ln \frac{\varepsilon
_2-\varepsilon _1}{\varepsilon _1}\right] ^{-1}\,\,.  \label{Tcx0}
\end{equation}
Note that, unless baths with negative temperatures are invoked, $\varepsilon
_2>2\varepsilon _1$ is needed.

While it is easy to analyz Eq. (\ref{Cy}) mathematically, it may be
helpful to provide an explicit example. We choose 
$\varepsilon _2=4\varepsilon _1$ and plot, in Fig. 1, $C_y/k_B$ 
against $T_y$ (in units $\varepsilon _1/k_B$) for various $T_x$'s. 
Note that, in this ``minimal'' system, $C_y$ does not change sign 
as long as $T_x$ is fixed.


\begin{figure}[tbp]
\par
\begin{center}
\begin{minipage}{.7\textwidth}
\hspace{2cm}
  \epsfxsize = .7\textwidth \epsfysize = .42\textwidth
  \epsfbox{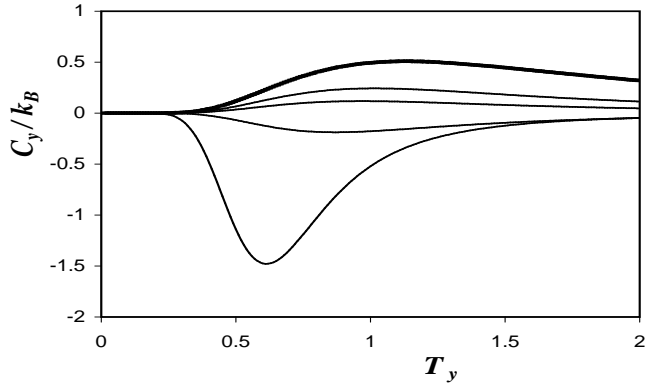}
\vspace{.2cm}
\caption{Specific heat $C_y/k_B$ vs. $T_y$ for various $T_x$'s.
Starting from the lowest curve, 
$k_BT_x / \varepsilon _1 = 1,2,3,4$ respectively.
The heavy line at the top is associated with $T_x = \infty $.} 
\end{minipage}
\end{center}
\end{figure}
%

\subsection{An intuitive picture}

To help the reader appreciate how such an unusual response arises, we
provide an intuitive picture for this phenomenon. Since the effects of a
thermal bath on the transition between two levels are written in terms of
rates, we may think of the role of a bath as that of (a pair of) ``pumps''
on the ``population'' in these levels. If we had only two levels, there
could be only one pair of rates (pumps), regardless of how many ``baths''
are coupled to them. In the steady state, only one ratio of populations
(say, the occupancy of the higher level to that of the lower) is relevant.
In this situation, the only sensible way to define ``thermodynamics'' of the
system is to have a {\em single} bath with an effective temperature. Then,
the usual intuitive picture easily emerges: if this (effective) temperature
is increased, the relative occupancy of the higher level will be higher,
leading to a positive specific heat. The steady state system might as well
be described as ``in equilibrium.'' Therefore, a minimal system which can
display negative specific heat must involve more than two (distinct) levels.

With three levels, there are three (sets of) rates, so that our system may
be coupled to three different ``thermal baths.'' By forbidding the
transition between the lower pair of levels (as in the simple case above),
it is already possible to generate ``unconventional'' population changes.
This restriction also reduces our consideration to only {\em two }baths.
Now, let us assume that our system has settled in a steady state where the
populations are given by $\left( P_0,P_1,P_2\right) $, using the notation
above. If we increase the temperature of the bath coupled to $1
\leftrightarrow 2$, we will certainly deplete $P_1$ in favor of $P_2$.
However, the $0 \leftrightarrow 2$ transition is coupled to a {\em different}
bath, with a fixed temperature, so that any increase in $P_2$ will be driven
to increases in $P_0$. If this bath is very cold, then the downward pump is
very strong (relative to the ``up-pump''), ensuring increases in $P_2$ be
accompanied by much larger increases in $P_0$. Thus, there is an effective
``downward'' shift of population (from $P_1$ to $P_0$), even though the
thermostat coupled to $1 \leftrightarrow 2$ has been turned up! By contrast,
if both transitions are coupled to the {\em same} bath, then there would be
also an ``upward'' shift (from $P_0$ to $P_2$), leading to the expected
increase in $U$. From these considerations, we see that forbidding the $1
\leftrightarrow 0$ transition introduces an ``barrier'' (or
``obstruction''), leading to a ``local minimum''. The presence of such
``obstructions'' forms the key ingredient to the display of negative
specific heats. In the next section, we provide another illustration of such
a link, in the context of currents and drives. Of course, ``the proof is in
the pudding.'' The heuristic argument, hopefully, helps the reader to digest
the mathematical ``pudding'' above.

Before turning to the explicit example, we note that this simple three state
system can be generalized to include a third bath, coupling to the $%
0\leftrightarrow 1$ transition. To be specific, let the temperature be $T_w$
and, in addition, let the ``efficiency'' of this coupling be $\eta $, so
that the extra rates are 
\begin{equation}
R(0\rightarrow 1)=\eta w\,;\quad R(1\rightarrow 0)=\eta 
\end{equation}
with 
\begin{equation}
w\equiv \exp \left[ -\varepsilon _1/k_BT_w\right] \,\,.
\end{equation}
Of course, setting $\eta $ to zero will reduce this system to the simple one
above. Computing the steady state distribution and the average internal
energy is a straightforward exercise. Here, we only quote the relevant
result. For sufficiently small $\eta $, there is a critical line in the $%
\left( T_x\text{-}T_w\right) $ plane, given by 
\begin{equation}
x_c+\eta w_c=\frac{\varepsilon _1-\eta \varepsilon _2}{\varepsilon
_2-\varepsilon _1}\,\,,
\end{equation}
on which the specific heat $C_y$ vanishes. For temperatures (or $x$,$w$)
below this line, $C_y$ is negative.

\subsection{An explicit example: the Driven Lattice Gas}

A ``realization'' of the above three-level system is the ``two-temperature
Ising model,'' \cite{KETT} which consists of a lattice gas coupled to two
baths. The only extra complication comes from degeneracies. Otherwise, all
analysis is identical to the minimal case.

Our system consists of a $2\times 3$ lattice, half filled with particles.
Imposing the boundary conditions: 
\[
\text{brick wall in }x\text{ ; periodic in }y 
\]
the 20 possible configurations fall into three distinct classes, labeled as
follows:\ \smallskip 

\begin{center}
\begin{tabular}{|c|c|}
\hline
$\,\bullet \,$ & \ \thinspace \ \  \\ \hline
$\,\bullet \,$ &  \\ \hline
$\,\bullet \,$ &  \\ \hline
\end{tabular}
\quad \quad \quad 
\begin{tabular}{|c|c|}
\hline
$\,\bullet \,$ &  \\ \hline
$\,\bullet \,$ & $\,\bullet \,$ \\ \hline
&  \\ \hline
\end{tabular}
\quad \quad \quad 
\begin{tabular}{|c|c|}
\hline
$\,\bullet \,$ &  \\ \hline
$\,\bullet \,$ &  \\ \hline
& $\,\bullet \,$ \\ \hline
\end{tabular}
\smallskip\ \smallskip\ 

0\quad \quad \quad \thinspace \thinspace \quad \quad 1\quad \quad \quad
\thinspace \thinspace \quad \quad 2\thinspace \thinspace \thinspace
\thinspace

\smallskip 
\end{center}

For energetics, we use the usual ferromagnetic Ising Hamiltonian: $-J$ is
associated with each nearest neighbor particle pair. For convenience, shift
all energies by $-3J$. As a result, the labels of the configurations are
also the energy levels (in units of $J$). The degeneracies are,
respectively, $\left( 2,12,6\right) $ due to translation and parity (in $x$
and $y$). To be pedantic, let us use a second label to distinguish them, so
that all 20 configurations are, explicitly:\smallskip\smallskip 

\begin{center}
\begin{tabular}{|c|c|}
\hline
$\,\bullet \,$ & \ \thinspace \ \  \\ \hline
$\,\bullet \,$ &  \\ \hline
$\,\bullet \,$ &  \\ \hline
\end{tabular}
\quad \quad \quad 
\begin{tabular}{|c|c|}
\hline
\ \thinspace \ \  & $\,\bullet \,$ \\ \hline
& $\,\bullet \,$ \\ \hline
& $\,\bullet \,$ \\ \hline
\end{tabular}
\smallskip\ \smallskip\ 

0,1\quad \quad \thinspace \thinspace \quad \quad \thinspace 0,2\thinspace
\thinspace \thinspace \thinspace \thinspace

\bigskip\ 

\begin{tabular}{|c|c|}
\hline
$\,\bullet \,$ &  \\ \hline
$\,\bullet \,$ & $\,\bullet \,$ \\ \hline
&  \\ \hline
\end{tabular}
\begin{tabular}{|c|c|}
\hline
$\,\bullet \,$ & $\,\bullet \,$ \\ \hline
$\,\bullet \,$ &  \\ \hline
&  \\ \hline
\end{tabular}
\begin{tabular}{|c|c|}
\hline
&  \\ \hline
$\,\bullet \,$ &  \\ \hline
$\,\bullet \,$ & $\,\bullet \,$ \\ \hline
\end{tabular}
\begin{tabular}{|c|c|}
\hline
&  \\ \hline
$\,\bullet \,$ & $\,\bullet \,$ \\ \hline
$\,\bullet \,$ &  \\ \hline
\end{tabular}
\begin{tabular}{|c|c|}
\hline
$\,\bullet \,$ & $\,\bullet \,$ \\ \hline
&  \\ \hline
$\,\bullet \,$ &  \\ \hline
\end{tabular}
\begin{tabular}{|c|c|}
\hline
$\,\bullet \,$ &  \\ \hline
&  \\ \hline
$\,\bullet \,$ & $\,\bullet \,$ \\ \hline
\end{tabular}
\begin{tabular}{|c|c|}
\hline
& $\,\bullet \,$ \\ \hline
$\,\bullet \,$ & $\,\bullet \,$ \\ \hline
&  \\ \hline
\end{tabular}
\begin{tabular}{|c|c|}
\hline
$\,\bullet \,$ & $\,\bullet \,$ \\ \hline
& $\,\bullet \,$ \\ \hline
&  \\ \hline
\end{tabular}
\begin{tabular}{|c|c|}
\hline
&  \\ \hline
& $\,\bullet \,$ \\ \hline
$\,\bullet \,$ & $\,\bullet \,$ \\ \hline
\end{tabular}
\begin{tabular}{|c|c|}
\hline
&  \\ \hline
$\,\bullet \,$ & $\,\bullet \,$ \\ \hline
& $\,\bullet \,$ \\ \hline
\end{tabular}
\begin{tabular}{|c|c|}
\hline
$\,\bullet \,$ & $\,\bullet \,$ \\ \hline
&  \\ \hline
& $\,\bullet \,$ \\ \hline
\end{tabular}
\begin{tabular}{|c|c|}
\hline
& $\,\bullet \,$ \\ \hline
&  \\ \hline
$\,\bullet \,$ & $\,\bullet \,$ \\ \hline
\end{tabular}
\smallskip\smallskip\ 

1,1\thinspace \thinspace \quad \thinspace 1,2\thinspace \thinspace \quad
\thinspace 1,3\thinspace \thinspace \quad \thinspace 1,4\thinspace
\thinspace \quad \thinspace 1,5\thinspace \thinspace \quad \thinspace
1,6\thinspace \thinspace \quad \thinspace 1,7\thinspace \thinspace \quad
\thinspace 1,8\thinspace \thinspace \quad \thinspace 1,9\thinspace \quad
\thinspace 1,10\quad \thinspace 1,11\quad \thinspace 1,12

\bigskip\ 

\begin{tabular}{|c|c|}
\hline
$\,\bullet \,$ &  \\ \hline
$\,\bullet \,$ &  \\ \hline
& $\,\bullet \,$ \\ \hline
\end{tabular}
\quad 
\begin{tabular}{|c|c|}
\hline
& $\,\bullet \,$ \\ \hline
$\,\bullet \,$ &  \\ \hline
$\,\bullet \,$ &  \\ \hline
\end{tabular}
\quad 
\begin{tabular}{|c|c|}
\hline
$\,\bullet \,$ &  \\ \hline
& $\,\bullet \,$ \\ \hline
$\,\bullet \,$ &  \\ \hline
\end{tabular}
\quad 
\begin{tabular}{|c|c|}
\hline
& $\,\bullet \,$ \\ \hline
& $\,\bullet \,$ \\ \hline
$\,\bullet \,$ &  \\ \hline
\end{tabular}
\quad 
\begin{tabular}{|c|c|}
\hline
$\,\bullet \,$ &  \\ \hline
& $\,\bullet \,$ \\ \hline
& $\,\bullet \,$ \\ \hline
\end{tabular}
\quad 
\begin{tabular}{|c|c|}
\hline
& $\,\bullet \,$ \\ \hline
$\,\bullet \,$ &  \\ \hline
& $\,\bullet \,$ \\ \hline
\end{tabular}
\smallskip\ \smallskip\ 

\quad \quad 2,1\thinspace \thinspace \thinspace \quad \quad 2,2\thinspace
\thinspace \thinspace \quad \quad 2,3\thinspace \thinspace \thinspace
\thinspace \quad \quad 2,4\thinspace \thinspace \thinspace \quad \quad
2,5\thinspace \thinspace \thinspace \quad \quad 2,6\smallskip\smallskip%
\qquad \quad \quad \quad
\end{center}

Next, let us specify the dynamics: particles may hop to nearest neighbor
empty sites (Kawasaki exchange \cite{Kawa}) according to the usual
Metropolis rates (Eq. \ref{metro}). The {\em two temperature} model is
defined so that exchanges along ($x,y$) axes are coupled to baths with
temperatures ($T_x,T_y$). As examples, the rate for 0,1$\rightarrow $2,1 is $%
\exp (-2J/kT_x)$ but the rate for 1,1$\rightarrow $2,1 is $\exp (-J/kT_y)$.
Obviously, our system is reduced to the trivial equilibrium case when $%
T_x=T_y$. With these rates, the master equations can be easily written.
Using subscripts to denote configurations, we give only three of the 20
master equations: 
\begin{eqnarray}
\partial _tP_{0,1} &=&P_{2,1}+P_{2,2}+P_{2,3}-3xP_{0,1}  \nonumber \\
\partial _tP_{1,1} &=&P_{2,1}+P_{2,3}+P_{1,2}+P_{1,7}+P_{1,4}-\left(
2y+3\right) P_{1,1} \\
\partial _tP_{2,1} &=&xP_{0,1}+y\left(
P_{1,1}+P_{1,2}+P_{1,3}+P_{1,6}\right) +P_{2,5}+P_{2,6}-7P_{2,1}  \nonumber
\end{eqnarray}
where 
\begin{equation}
x\equiv \exp (-2J/kT_x)\,\,,\,\,y\equiv \exp (-J/kT_y)\,\,.
\end{equation}
To find the steady state $P^{*}$'s, we note that all degenerate
configurations will have same probability (since they are related to each
other by symmetries). So, we just write equations without the second
subscript: 
\begin{eqnarray}
0 &=&3P_2^{*}-3xP_0^{*}  \nonumber \\
0 &=&2P_2^{*}-2yP_1^{*} \\
0 &=&xP_0^{*}+4yP_1^{*}-5P_2^{*}  \nonumber
\end{eqnarray}
Apart from degeneracies, the solutions are same as in the minimal example
above: 
\begin{equation}
P_0^{*}=y/Z\,;\,P_1^{*}=xZ\,;\,\,P_2^{*}=xyZ\,\,,
\end{equation}
with 
\begin{equation}
Z=2y+12x+6xy\,\,.
\end{equation}
The average energy is given by 
\begin{equation}
\frac UJ=\frac{12x+2\times 6xy}Z=6x\frac{1+y}{y+6x+3xy}\,\,,
\end{equation}
so that 
\begin{equation}
C_y\equiv \frac{\partial U}{\partial T_y}\propto \left[ 3x-1\right] .
\end{equation}
(with a positive definite proportionality factor). Similar to the minimal
case, we have 
\begin{equation}
C_y<0
\end{equation}
for all $T_y$, provided $T_x$ is low enough. Note that the degeneracies of
these three levels affect the critical $T_{xc}$. A simple exercise including 
$g_\alpha $ (the degeneracy of level $\alpha $) in the computations above
leads to the general result 
\begin{equation}
T_{xc}\equiv \frac{\varepsilon _2}{k_B}\left[ \ln \frac{\varepsilon
_2-\varepsilon _1}{\varepsilon _1}+\ln \frac{g_2}{g_0}\right] ^{-1}\,\,.
\label{TcxGen}
\end{equation}
Interestingly, this critical value is independent of the degeneracy of the
middle level. From the intuitive picture given above, this behavior is,
arguably, reasonable. After all, the transition from positive to negative
specific heat depends how the depopulation of the middle level is
redistributed between the upper and lower levels. At the critical point,
this redistribution leads to an average energy (associated with the outer
levels alone) which is identical to $\varepsilon _1$, leading to {\em no}
change in the total energy. Now, each state in the middle level would
``suffer the same fate,'' so that it is irrelevant how many such states
there are. Thus, we have $g_1$ independence.

\subsection{Energy flux}

In general, a system coupled to two thermal baths will lead to a transfer of
energy from the hotter bath to the colder one. On the other hand, by
definition of steady state, the energy stored in our system is a constant.
Thus, we may expect a constant energy flux {\em through} our system. Most
remarkably, in our minimal ($\eta =0$) case, there is {\em no} energy flux
through the system! In this sense, our system and both baths are ``in
equilibrium,'' and detailed balance is satisfied! Indeed, the general
condition for detailed balance \cite{DM} is less restrictive than Eq. (\ref
{DB1}), so that it is possible to have a system coupled to two baths {\em and%
} retain detailed balance. The ramifications of such ``effectively
equilibrium'' systems remain to be explored.

Meanwhile, as $T_y$ is increased (with fixed $T_x<T_{xc}$), the energy of
the system decreases. To be more precise, suppose we take a system in steady
state and suddenly replace the $y$-bath by another with a {\em higher}
temperature. Then there must be a net flow of energy {\em out of} our system
somehow, since the new steady state must have {\em lower }$U$. This paradox
cannot be solved by studying the properties of steady states alone. The
resolution lies in an analysis of the full dynamics. Deferring all details
to another article, we provide only the result here: less energy is
transferred from the $y$-bath to our system than from the system to the $x$%
-bath. In the general case ($\eta >0$), there is a net energy flux through
our system, so that similar surprising behavior appears less paradoxical.

\section{A Model with Negative Conductivity}

The counter-intuitive phenomenon of negative response is not restricted to
the energy-temperature variables. In this section, we show a similar
behavior in the current-drive variables, i.e., the possibility of ``negative
conductivity''.

\subsection{Set up and the unobstructed case}

Consider a {\em single} particle hopping in two ``lanes'' of $L$ sites,
periodic on the long side and ``brick walls'' bounding the lanes. For
convenience, assume $L$ is even, so that the sites along the lanes can be
labeled by the integer $i=-L/2,\ldots ,L/2-1$. Periodicity allows us to use $%
i=0,1,\ldots ,L-1$ interchangeably.

Let us denote the probabilities of finding the particle on the two lanes by 
\begin{equation}
P_i\,\,\,\,\text{and}\,\,\,Q_i\,\,.
\end{equation}
Symbolically, we represent this system by:\smallskip

\begin{equation}
\cdots 
\begin{tabular}{|c|c|c|c|c|}
\hline
$Q_{-2}$ & $Q_{-1}$ & $\,\,Q_0\,\,$ & $\,\,Q_1\,\,$ & $\,\,Q_2\,\,$ \\ \hline
$P_{-2}$ & $P_{-1}$ & $P_0$ & $P_1$ & $P_2$ \\ \hline
\end{tabular}
\cdots  \label{PQ}
\end{equation}

Let there be no obstructions and suppose the particle is ``driven'' by a
uniform external ``electric'' field $E$, so that the jump rates are
proportional to 
\begin{eqnarray}
1\quad  &&\text{for crossing lanes only;}  \nonumber \\
x\quad  &&\text{for jumping ``upstream'' (}i\rightarrow i-1\text{) to {\em %
either} lane;} \\
1/x\quad  &&\text{for jumps ``downstream'' (}i\rightarrow i+1\text{) to {\em %
either} lane.}  \nonumber
\end{eqnarray}
The normalization factor is just 
\begin{equation}
N=\frac 1{1+2x+2/x}\,\,.
\end{equation}
We may regard $x$ as a Boltzmann-like rate: 
\begin{equation}
x=e^{-\beta E}\,,
\end{equation}
in which case $N=1/(1+4\cosh \beta E)$. Note that the behavior of this
``charged'' particle is not the same as one in free space. Instead, it
represents an overdamped situation, where inertia can be neglected compared
to (thermal) damping. As a result, the particle settles down to the drift
velocity instantaneously. Furthermore, with the rates and normalization
chosen, the limit of $E\rightarrow \infty $ does {\em not} mean the particle
will move with infinity velocity. From the rates, it should be clear that,
in this limit, all that happens is that the particle moves to the next site
(either lane) with unit probability. Thus, the {\em velocity saturates} at
unity for ``infinite drive''.

Without obstructions, the system is translationally invariant and the steady
state is trivially given by 
\begin{equation}
P_i=Q_i=1/2L
\end{equation}
since the particle can be located at any of the $2L$ locations. The uniform
current density, $J_{free}$, can be found by considering the probability of
all jumps between $i$ and $i+1$. Taking into account the contributions from
both lanes, we have $\frac 2xN\left( P_i+Q_i\right) -2xN\left(
P_{i+1}+Q_{i+1}\right) $. Therefore we obtain 
\begin{equation}
J_{free}=\frac{2\sinh \beta E}{(1+4\cosh \beta E)L}\,\,.
\end{equation}
Needless to say, the conductivity is positive 
\begin{equation}
\frac{\partial J}{\partial E}>0
\end{equation}
for all finite $\beta E$.

Another perspective is to consider the average velocity of the particle.
Since we have only a single particle, we have $J=\rho v$. Given uniform
density ($1/2L$), we obtain $v=4\sinh \beta E/(1+4\cosh \beta E)$, which is
a monotonically increasing function of the drive $E$. Note that this
expression is expected for both small and large $E.$ In the former case,
four out of the five possible jumps contribute to the velocity, so that the
conductivity is simply $(4/5)\beta $.

\subsection{Single obstruction case}

Next, let us introduce barriers at $i=0$, but only for {\em one lane.}
Specifically, ``impenetrable walls'' are placed between the site associated
with $P_0$ and three of its neighbors (those associated with $Q_0$, $P_1$,
and $Q_1$). Symbolically, we draw thick lines around $P_0$ (Fig. 2), and,
using the analogy of water flowing under gravity, we will refer to the
obstruction as a ``cup.'' Clearly, the cup causes a blockage, which, for
large $E$, will reduce the current seriously. However, as $E$ is lowered,
the particle's chances of backward jumps are higher and leads to a {\em %
higher} current, giving us a{\em \ negative }conductivity.


\begin{figure}[tbp]
\par
\begin{center}
\vspace{-.5cm}
\begin{minipage}{.7\textwidth}
\hspace{1cm}
  \epsfxsize = .75\textwidth \epsfysize = .21\textwidth 
  \epsfbox{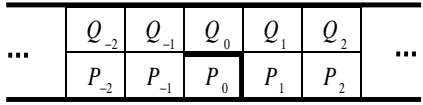}
\caption{Schematic display of the probability for finding the particle 
in each box. Thick lines borders are associated with impenetrable walls.}
\end{minipage}
\end{center}
\end{figure}
%

To see that this picture is indeed correct, we start with a careful
consideration of all the master equations. Without the cup, these equations
read, for all $i$, 
\begin{eqnarray}
\partial _tP_i &=&\frac 1x\left( P_{i-1}+Q_{i-1}\right) +x\left(
P_{i+1}+Q_{i+1}\right) +Q_i-\left( 1+\frac 2x+2x\right) P_i  \label{MEfree}
\\
\partial _tQ_i &=&\frac 1x\left( P_{i-1}+Q_{i-1}\right) +x\left(
P_{i+1}+Q_{i+1}\right) +P_i-\left( 1+\frac 2x+2x\right) Q_i  \nonumber
\end{eqnarray}
With the cup in place, the only jumps affected are those involving $%
0\leftrightarrow 1$. As a result, the only equations affected are: 
\begin{eqnarray}
\partial _tP_0 &=&\frac 1x\left( P_{-1}+Q_{-1}\right) -2xP_0  \label{MEob} \\
\partial _tQ_0 &=&\frac 1x\left( P_{-1}+Q_{-1}\right) +x\left(
P_1+Q_1\right) -2\left( \frac 1x+x\right) Q_0  \nonumber \\
\partial _tP_1 &=&\frac 1xQ_0+x\left( P_2+Q_2\right) +Q_1-\left( 1+\frac 2x%
+x\right) P_1  \nonumber \\
\partial _tQ_1 &=&\frac 1xQ_0+x\left( P_2+Q_2\right) +P_1-\left( 1+\frac 2x%
+x\right) Q_1  \nonumber
\end{eqnarray}
A factor of $N$ has been absorbed into the time scale, an irrelevant concern
for the time-independent state. Setting the left hand side of these
equations to zero, we seek the unique solution corresponding to the steady
state. Note that, if we set the drive to zero ($x\rightarrow 1$), then the
obstruction becomes irrelevant in steady state and we retrieve the expected
result: the trivial flat distribution. For the $E\neq 0$ case, we defer all
details of the calculation to the appendix, providing a few brief steps and
the result here.

For convenience, define the sums and differences: 
\begin{eqnarray}
S &\equiv &P+Q \\
D &\equiv &P-Q
\end{eqnarray}
With no obstructions, it is intuitively clear (and mathematically easy to
show) that the two lanes have equal probabilities in the steady state.
However, the cup enhances $P_0$, so that 
\begin{eqnarray}
D_i &=&0\quad \text{for }i\neq 0  \nonumber \\
D_0 &=&\frac{1-x^2}{1+2x^2}S_0>0  \label{D0}
\end{eqnarray}
Invoking (probability) current conservation, we set 
\begin{equation}
K=\frac 2xS_{i-1}-2xS_i\quad i\neq 0,1
\end{equation}
which is proportional to the current between $i-1$ and $i$. But this $K$
must also hold for {\em all} links. In particular, 
\begin{eqnarray}
K &=&\frac 2xS_{-1}-2xS_0 \\
K &=&\frac 2xQ_0-xS_1=\frac{S_0-D_0}x-xS_1  \nonumber
\end{eqnarray}
Using Eq. (\ref{D0}), we find a set of equations for the $S$'s, leading to

\begin{equation}
S_0=\frac{\left( 1+2x^2\right) \left( 2-x^2-x^{2L}\right) }{%
3Lx^2+2+\allowbreak 2x^2-Lx^{2L}-2Lx^{2+2L}-2x^{2L}-2x^{2+2L}}  \label{S0}
\end{equation}
and 
\begin{equation}
K=\frac{2x\left( 1-x^2\right) \left( 3-x^{2\left( L-1\right)
}-2x^{2L}\right) }{3Lx^2-Lx^{2L}-2Lx^{2+2L}-2x^{2L}+2-2x^{2+2L}+\allowbreak
2x^2}  \label{K}
\end{equation}
It is instructive to check that these expressions indeed reduce to $1/L$ and 
$0$, respectively, in the limit $E\rightarrow 0$ ($x\rightarrow 1$). Another
interesting limit is $\beta E$ diverges faster than $\ln L$ ($Lx\rightarrow 0
$). In this case, the probability of the particle jumping out of the cup is
so low that even the entropy factor (the $2L$ possible locations for the
particle) is not enough to overcome the barrier ($x\rightarrow 0$), so that
``complete trapping'' occurs. This expectation is borne out by 
\begin{equation}
\left[ S_0\right] _{trapped}\rightarrow 1
\end{equation}
and, from Eq. (\ref{D0}), $\left[ D_0\right] _{trapped}\rightarrow 1$ also.
Thus, we have 
\begin{equation}
\left[ P_0\right] _{trapped}\rightarrow 1\quad \text{and}\quad \left[
Q_0\right] _{trapped}\rightarrow 0\,.
\end{equation}
Verifying that all other probabilities also vanishes, we see that the
particle is completely trapped in the cup. To complete the study of this
limit, we find 
\begin{equation}
K_{trapped}=3x+O(Lx^2,x^2)\rightarrow 0
\end{equation}
Since $J=NK$, the current {\em vanishes} (as $x^2$) for large fields, rather
than saturating, as in the free case. Thus, $\partial J/\partial E$ must be
negative in some range of $E$.

Returning to Eqs. (\ref{S0}) and (\ref{K}), we find less cumbersome forms if
we consider the thermodynamic limit and drop terms of $O(x^{2L},Lx^{2L})$
first. The results are:

\begin{eqnarray}
S_0 &=&\frac{\left( 1+2x^2\right) \left( 2-x^2\right) }{3Lx^2+2+\allowbreak
2x^2} \\
J_{obstruction} &=&NK=\frac{6x^2\left( 1-x^2\right) }{\left( 2+\allowbreak
2x^2+3Lx^2\right) \left( 2+x+2x^2\right) }
\end{eqnarray}


\begin{figure}[tbp]
\par
\begin{center}
\begin{minipage}{.7\textwidth}
\hspace{2cm}
  \epsfxsize = .7\textwidth \epsfysize = .42\textwidth
  \epsfbox{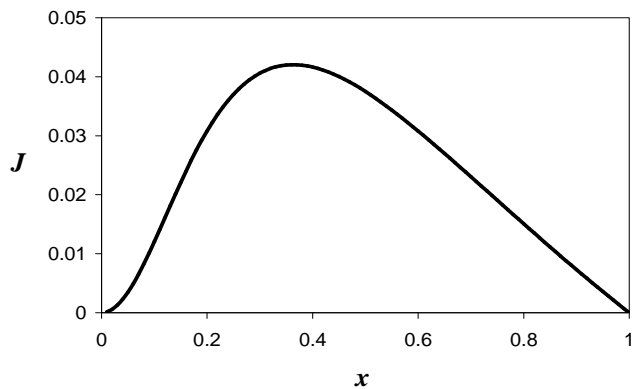}
\vspace{.2cm}
\caption{Current vs. $x \equiv e^{-\beta E}$. 
Note increasing $E$ corresponds to decreasing $x$ and the linear rise of 
the current for small $E$ (near $x = 1$). For large enough drive (small $x$),
the current drops with increasing $E$.}
\end{minipage}
\end{center}
\end{figure}
%

Of course, we can differentiate the current and find the critical field
above which the conductivity $\partial J/\partial E$ is negative. However,
it may be more instructive to provide an illustration of this expression. In
Fig. 3, we plotted this current as a function of the driving field, for the
case of $L=10$. We should remind the reader that $x=\exp (-\beta E)$, so
that larger $x$ corresponds to smaller $E$. Note that the curve is basically
linear near the $x=1$ end, confirming the typical constant, positive
conductivity for small $E$. On the other hand, {\em negative conductivity}
sets in for sufficiently large $E$ (approximately $k_BT$ in this case).

\section{Concluding Remarks}

In this paper, we have shown that ``negative responses,'' in a system
coupled to thermal baths, can be easily generated, provided we consider {\em %
non}-equilibrium steady states. Specifically, we gave examples of
exceedingly simple systems which display negative specific heat. One is an
abstract system with only three energy levels, but coupled to {\em two}
thermal baths. The other is a standard Ising lattice gas ($3$ particles in $%
2\times 3$ sites), also coupled to two baths. These models are ``minimal''
in the sense that systems with just two energy levels can support only a
single (set of) transition rates, so that an single (effective) temperature
can be defined. Then, we have essentially an equilibrium system, with
non-negative specific heats. There is no doubt that such systems can be
extended to more complex and macroscopic cases. In particular, using Monte
Carlo simulation techniques with the two-temperature Ising model, we have
observed negative specific heats in, e.g., a $60\times 60$ lattice for $%
T_y\gtrsim 1.5T_O$ with $T_x=0.7T_O$. Another example is a two-temperature
Ising model with a population of mobile impurities that couple to a heat
bath with a temperature different from that of the Ising spins. This model
has been proposed to mimic the non-equilibrium behavior of a biological
membrane in which diffusing protein particles receive energy from a heat
bath with a temperature that is different from the membrane matrix in which
diffuse. Under certain conditions this has shown to lead to a negative
specific heat (Fig. 4 in \cite{OGM}). We also showed a second form of
negative response: negative conductivity. The explicit example consists of
particles hopping along two lanes of discrete sites, driven by an external
field so that a non-vanishing DC current is established in the steady state.
When an obstruction is introduced, the current is shown to decrease when the
drive is increased (beyond some critical value). In both cases, the key
ingredient (besides being in {\em non}-equilibrium steady states) is the
presence of local minima or ``obstructions.'' When the drive, be it another
thermal bath or an external field, is too large, the system is caught for
longer in the obstruction. By lowering the drive, the system can respond
``more positively''. Though the phenomenon has been observed for some time,
e.g., in driven diffusive systems \cite{KLS}, it has been discovered in
another context and popularized through the catchy phrase: ``{\em freezing
by heating}'' \cite{FbH}. Since this kind of counter-intuitive response is
not only reserved for negative specific heat, but also negative conductivity
and beyond, we may label all of them by the catch-all phrase: ``{\em getting
more by pushing less.}''

Though we are certain that this kind of phenomena are present in physical
systems, we have not conducted a systematic search for them, nor have we
proposed designs for such devices. The goal of this paper is simply to
encourage those who teach statistical mechanics to keep their students'
minds open about the issue of negative specific heats. In particular, in
non-equilibrium steady states, positive responses are {\em not }intimately
related to stability.

To close, we should mention that ``negative specific heat'' is not a novel
phrase. Indeed, it occurs frequently in discussions of self-gravitating
systems. Evidently, Eddington wrote about how a star or star-cluster would
cool down if energy is added \cite{Eddington}. Reading the literature, there
is no doubt that this is a subtle situation of a complex system, with
particles subjected to long range interactions (gravity). Furthermore, it is
clear that such systems are inherently unstable, leading to ``gravothermal
catastrophe'' and collapse. Clearly, the theoretical existence of such a
kind of negative specific heat was not well received by physicists. In the
opening sentence of a recent article \cite{LB}, Lynden-Bell wrote, ``When I
first used the concept of Negative Specific Heat ...the Statistical
Mechanics community thought I was talking nonsense.'' By contrast, our
examples are far less exotic and the phenomenon should be abundant in our
neighborhood as well as amongst the distant stars. After all, every living
organism can be considered as a non-equilibrium steady state, coupled to
more than one reservoir of energy so that there is a constant through-flux.
If we succeed in convincing other teachers to add a footnote in their
course, that $\partial U/\partial T\geq 0$ is ironclad {\em only} for system
in equilibrium, then this paper would have served its purpose.

\section{Appendix}

Here we provide a few more details for finding the steady state distribution
in the two-lane hopping model. Referring to the schematic diagram (\ref{PQ}%
), we seek solutions to the set of equations (\ref{MEfree},\ref{MEob}) with
zero on the right hand sides. Recall that we have periodic boundary
conditions, so that hopping rates from sties $i=L-1$ to $i=-L$ are the same
as all others (except for the obstruction).

As mentioned in the text, it is convenient to define sums and differences: 
\begin{eqnarray}
S &\equiv &P+Q  \nonumber \\
D &\equiv &P-Q
\end{eqnarray}
so that we have 
\begin{eqnarray}
0 &=&\frac 2xS_{i-1}+2xS_{i+1}-2\left( \frac 1x+x\right) S_i \\
0 &=&-2\left( 1+\frac 1x+x\right) D_i
\end{eqnarray}
and 
\begin{eqnarray}
0 &=&\frac 2xS_{-1}+xS_1-2xS_0-\frac 1x\left( S_0-D_0\right)  \\
0 &=&-xS_1-2xD_0+\frac 1x\left( S_0-D_0\right)  \\
0 &=&\frac 1x\left( S_0-D_0\right) +2xS_2-\left( \frac 2x+x\right) S_1 \\
0 &=&-\left( 2+\frac 2x+x\right) D_1
\end{eqnarray}
from Eqs. (\ref{MEfree},\ref{MEob}), respectively. The advantage of this
decomposition is obvious now. It proves the intuitive notion that, apart
from the obstructed sites, the two lanes should have equal probabilities in
the steady state. So, 
\begin{eqnarray}
D_i &=&0\quad \text{for }i\neq 0 \\
D_0 &=&\frac{1-x^2}{1+2x^2}S_0
\end{eqnarray}
Now, we have a set of equations involving only $S$: 
\begin{eqnarray}
0 &=&\frac 2xS_{-1}+xS_1-\frac{5+4x^2}{1+2x^2}xS_0  \label{S00} \\
0 &=&\frac{3x}{1+2x^2}S_0+2xS_2-\left( \frac 2x+x\right) S_1  \label{S1} \\
0 &=&\frac 2xS_{i-1}+2xS_{i+1}-2\left( \frac 1x+x\right) S_i\quad \text{for}%
\quad i\neq 0,1  \label{S}
\end{eqnarray}

To solve these, we exploit current conservation, since the physical content
of these equations lies in the difference between the particle currents into
and out-of site $i$ must be zero. Thus, we first consider the ``unaffected''
sites, just to be careful. Set $2S_{i-1}/x-2xS_i$ to a constant: 
\begin{equation}
K=\frac 2xS_{i-1}-2xS_i\quad i\neq 0,1
\end{equation}
which will be proportional to the steady state current. (As a reminder,
given the periodic condition, $S_{L/2}\equiv S_{-L/2}$, $%
K=2S_{L/2-1}-2xS_{-L/2}$ is part of this set of equations.) Note that this
``trick'' solves Eq. (\ref{S}) automatically: $0=K-K$. Next, this $K$ serves
as {\em the} steady state current, so that it must hold for {\em all} jumps $%
i\leftrightarrow i+1$. In particular, the jumps across $-1\leftrightarrow 0$
leads to 
\begin{equation}
K=\frac 2xS_{-1}-2xS_0
\end{equation}
which is the same as the unaffected case. Meanwhile, the jumps across $%
0\leftrightarrow 1$ provides a new equation: 
\begin{equation}
K=\frac 2xQ_0-xS_1
\end{equation}
By definition, $2Q_0=S_0-D_0$, so that, using Eq. (\ref{D0}), we obtain 
\begin{equation}
2Q_0=\frac{3x^2}{1+2x^2}S_0
\end{equation}
and 
\begin{equation}
K=\frac{3x}{1+2x^2}S_0-xS_1
\end{equation}
All these equations with $K$ on the left can be recast as recursion
relations for the $S_i$'s. Specifically, we will try to express everything
in terms of $S_0$, so that we write 
\begin{eqnarray}
S_1 &=&\frac 3{1+2x^2}S_0-\frac 1xK  \label{R0} \\
S_{i-1} &=&\frac x2K+x^2S_i\quad \quad \text{for}\quad \quad i\neq 1
\label{R1}
\end{eqnarray}
Starting with the $i=0$ recursion relation Eq. (\ref{R1}), we work all the
way around back to $S_1$: 
\begin{eqnarray}
S_{-1} &=&\frac x2K+x^2S_0 \\
S_{-2} &=&\frac x2K+x^2S_{-1}=\frac x2\left( 1+x^2\right) K+x^4S_0  \nonumber
\\
&&\vdots   \nonumber \\
S_{-n} &=&\frac x2\left( 1+\cdots +x^{2(n-1)}\right) K+x^{2n}S_0  \nonumber
\\
&=&\frac{x\left( 1-x^{2n}\right) }{2\left( 1-x^2\right) }K+x^{2n}S_0 
\nonumber \\
&&\vdots   \nonumber \\
S_1 &=&S_{-L+1}=\frac{x\left( 1-x^{2(L-1)}\right) }{2\left( 1-x^2\right) }%
K+x^{2(L-1)}S_0  \nonumber
\end{eqnarray}
Together with Eq. (\ref{R0}), the last of these allows us to express $K$ in
terms of $S_0$. 
\begin{equation}
\left[ \frac{x^2\left( 1-x^{2(L-1)}\right) }{2\left( 1-x^2\right) }+1\right]
K=x\left[ \frac 3{1+2x^2}-x^{2(L-1)}\right] S_0  \label{KS}
\end{equation}
Finally, normalization ($\sum_iS_i=1$) will fix everything.

It may be instructive to compare the above equations with those in the free
case. The only difference is that we would have one more ``unaffected''
step. Instead of Eq. (\ref{R0}), we have $\left[ S_0\right] _{free}=\frac x2%
K_{free}+x^2\left[ S_1\right] _{free}$, so that 
\begin{eqnarray}
x^{-2}\left[ S_0\right] _{free}-\frac 1{2x}K_{free}=\frac{x\left(
1-x^{2(L-1)}\right) }{2\left( 1-x^2\right) }K_{free}+x^{2(L-1)}\left[
S_0\right] _{free}
\end{eqnarray}
i.e., 
\begin{equation}
\left[ S_0\right] _{free}=\frac x{2\left( 1-x^2\right) }K_{free}
\end{equation}
Note first that this expression is {\em independent of }$L$! This result
should be expected, since we know that the average velocity must be $L$%
-independent for the free case. But the velocity is just the current divided
by the particle density (i.e., probability for finding the particle at, say,
site $0$). To continue the check, insert this expression into the recursion
relations and find that all the $S$'s are the same. So, normalization gives $%
\left[ S_i\right] _{free}=1/L$ and $K_{free}=\left( 4\sinh \beta E\right) /L$%
.

Armed with these considerations, we return to the case with barrier and
define 
\begin{equation}
M\equiv S_0-\frac x{2\left( 1-x^2\right) }K  \label{M}
\end{equation}
which is a measure of the effect of the cup. Eliminating $K$ in its favor,
Eq. (\ref{KS}) is replaced by a neater formula: 
\begin{equation}
\left[ \frac{\left( 1-x^{2L}\right) }{\left( 1-x^2\right) }+1\right] M=\frac{%
2\left( 1+x^2\right) }{\left( 1+2x^2\right) }S_0  \label{MS0}
\end{equation}
Finally, the sum over the $S$'s is

\begin{eqnarray}
S_0+\sum_{n=1}^{L-1}S_{-n} &=&S_0+\sum_{n=1}^{L-1}\left[ \frac{x\left(
1-x^{2n}\right) }{2\left( 1-x^2\right) }K+x^{2n}S_0\right]   \nonumber \\
&=&S_0+\frac{x(L-1)}{2\left( 1-x^2\right) }K+\sum_{n=1}^{L-1}x^{2n}M
\end{eqnarray}
Eliminating $K$ and setting this to unity, we have 
\begin{equation}
1=LS_0+\left[ \frac{1-x^{2L}}{1-x^2}-L\right] M  \label{MS1}
\end{equation}
Eqs. (\ref{MS0},\ref{MS1}) can now be used to obtain Eq. (\ref{S0}).

\begin{center}
{\bf {ACKNOWLEDGMENTS} }
\end{center}

This research is supported in part by grants from the Danish Natural Science
Research Council and the US National Science Foundation (DMR-9727574) . One
of us (RKPZ) thanks J.L. Lebowitz for illuminating correspondence and H.W.
Diehl for his hospitality at the University of Essen, where most of this
work was performed.

\end{document}